\documentclass{article}

\usepackage{microtype}
\usepackage{graphicx}
\usepackage{subfigure}
\usepackage{booktabs} 

\usepackage{hyperref}

\usepackage[accepted]{icml2025}

\usepackage{amsmath}
\usepackage{amssymb}
\usepackage{mathtools}
\usepackage{amsthm}

\usepackage[capitalize,noabbrev]{cleveref}

\theoremstyle{plain}

\theoremstyle{definition}

\theoremstyle{remark}

\usepackage[textsize=tiny]{todonotes}

\usepackage{lineno}

\usepackage{color}
\usepackage{xspace}
\usepackage{pifont} 
\newcommand{\cmark}{\textcolor{green}{\ding{51}}\xspace} 
\newcommand{\xmark}{\textcolor{red}{\ding{55}}\xspace} 

\definecolor{darkblue}{rgb}{0, 0, 0.5}
\hypersetup{colorlinks=true, citecolor=darkblue, linkcolor=darkblue, urlcolor=darkblue}

\icmltitlerunning{Investigating Tool-Memory Conflicts in Tool-Augmented LLMs}

\begin{document}

\twocolumn[
\icmltitle{Investigating Tool-Memory Conflicts in Tool-Augmented LLMs}



\icmlsetsymbol{equal}{*}

\begin{icmlauthorlist}
\icmlauthor{Jiali Cheng}{yyy}
\icmlauthor{Rui Pan}{comp}
\icmlauthor{Hadi Amiri}{yyy}
\end{icmlauthorlist}

\icmlaffiliation{yyy}{University of Massachusetts Lowell, USA}
\icmlaffiliation{comp}{University of Illinois Urbana-Champaign, USA}

\icmlcorrespondingauthor{Jiali Cheng}{jiali\_cheng@uml.edu}

\icmlkeywords{Knowledge Conflict, Tool-Augmented LLM}

\vskip 0.3in
]



\printAffiliationsAndNotice{}  

\begin{abstract}
Tool-augmented large language models (LLMs) have powered many applications. However, they are likely to suffer from knowledge conflict. In this paper, we propose a new type of knowledge conflict -- Tool-Memory Conflict (TMC), where the internal parametric knowledge contradicts with the external tool knowledge for tool-augmented LLMs.
We find that existing LLMs, though powerful, suffer from TMC, especially on STEM-related tasks. We also uncover that under different conditions, tool knowledge and parametric knowledge may be prioritized differently. We then evaluate existing conflict resolving techniques, including prompting-based and RAG-based methods. Results show that none of these approaches can effectively resolve tool-memory conflicts.
\end{abstract}
\section{Introduction}
Tool-augmented large language models (LLMs) are LLMs that can use external tools, such as function calling and APIs~\citep{schick2023toolformer,patil2023gorilla,qin2024tool}. By integrating external tools, the LLMs exhibit enhanced problem-solving capabilities, especially in applications that require interact with physical world and access knowledge bases. This augmentation extends their functional scope, enabling interaction with dynamic and domain-specific information sources~\citep{chen-etal-2023-chatcot,ma2024sciagent,cheng2025atlas,lu-etal-2025-tart}.

Despite the advantages, the integration of external tools introduces potential epistemic inconsistencies, as tool-generated outputs may contradict the parametric knowledge encoded within the parameters of LLMs~\citep{xu2024knowledge,cheng2025tool}. For instance, Temporal Discrepancy arises when external tools provide updated information that conflicts with the static, pre-trained knowledge of an LLM to a specific cutoff date. Additionally, mechanistic differences between internal LLM memory and external tools complicate the resolution of such inconsistencies, raising critical concerns regarding knowledge reliability and coherence.

While prior research has examined several types of knowledge conflicts in LLMs, including context-memory conflict and inter-context conflict, limited attention has been given to discrepancies arising between parametric memory and external tool outputs. This study introduces the concept of tool-memory conflict, a novel category of knowledge inconsistency in LLMs, wherein the internal parametric knowledge of the model diverges from the outputs of external tools when addressing the same query. Specifically, we aim to answer the following research questions:

\begin{itemize}
    \item RQ1: Under what conditions (task, scale of LLM) do tool-memory conflicts appear in tool-augmented LLMs?
    \item RQ2: When confronted with a tool-memory conflict, do LLMs prioritize parametric knowledge or tool-generated outputs?
    \item RQ3: What methodologies can effectively reconcile tool-memory conflicts across varying contexts?
\end{itemize}

By addressing these questions, this study seeks to advance the understanding of knowledge integration in tool-augmented LLMs and develop strategies for mitigating epistemic inconsistencies, ultimately enhancing the reliability and interpretability of tool-augmented language models.

Through experiments and analysis, we find that LLMs can have significant amount of tool-memory conflicts across a variety of tasks, including math, QA. Although LLMs are trained to incorporate external tools during tool-augmentation to complement their memory, especially on professional or time-sensitive tasks, they can still prioritize internal memory over tools.

The contributions of this paper are
\begin{itemize}
    \item formulating tool-memory conflict, a new type of knowledge conflict in LLMs, and discussing its importance, causes, and differences to existing knowledge conflicts;
    \item investigating how LLMs prioritize knowledge under tool-memory conflict, and the bias of LLMs; 
    \item evaluating a wide range of methods to resolve tool-memory conflict.
\end{itemize}


\section{Related work}
\paragraph{Knowledge Conflicts in LLMs}
Existing works focus on three types of knowledge conflicts~\citep{xu-etal-2024-knowledge-conflicts}, 1) context-memory conflict, where the output of internal parametric knowledge differs from the contextual knowledge~\citep{longpre-etal-2021-entity,tan-etal-2024-blinded,gekhman-etal-2023-trueteacher,wang2024resolving}, 2) inter-context conflict, where conflicts happens between various pieces of contextual information~\citep{zhang-choi-2023-mitigating,chen-etal-2022-rich,wan-etal-2024-evidence,jin-etal-2024-tug}, and 3) inter-memory conflict, where conflicts happen internally in the parametric knowledge of LLMs~\citep{lee2022factuality,wang-etal-2023-causal,qi-etal-2023-cross,hase-etal-2023-methods}.


\paragraph{Tool-Augmented LLMs}
Tool-Augmented LLMs are LLMs that know how to call external tools to answer questions the their internal memory is insufficient at~\citet{schick2023toolformer}. These LLMs are usually explicitly trained using tool using demonstrations that are synthesized with seed samples and LLMs~\citep{patil2023gorilla,li-etal-2023-api,tang2023toolalpaca}. modifying existing datasets~\citep{basu-etal-2024-api}, and dataset development with GPT-4~\citep{qin2024toolllm}.
TAML~\citep{parisi2022talm} used self-play to boost LLMs' performance on math and reasoning tasks.

\begin{figure*}
    \centering
    \includegraphics[width=0.98\linewidth]{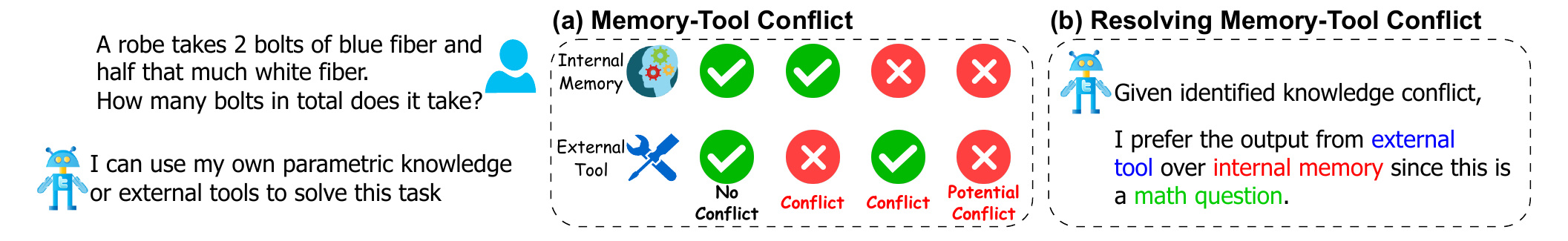}
    \caption{Illustration of tool-memory Conflict (MCT). \cmark and \xmark mean the output is correct or wrong respectively compared to the ground truth answer.}
    \label{fig:mct}
\end{figure*}




\section{Tool-Memory Conflict}\label{sec:prem}

\paragraph{Problem Definition}
Let $f$ denote a tool-augmented LLM capable of calling external tools from tool set $\mathcal{T}$. The \emph{tool-memory conflict} (TMC) occurs if the parametric knowledge of $f$ conflicts with the external tool knowledge of $\mathcal{T}$, more formally
\begin{equation}
    f(q) \neq f(q; \mathcal{T}),
\end{equation} where $q$ is a query.

\subsection{Examples of tool-memory Conflict}
There exist numerous scenarios where conflicts may arise between the internal memory of LLMs and external tools.

\paragraph{Example 1: Mathematics Problem}
When posed with a mathematical question, an LLM may respond using its generative capabilities by recalling patterns from its training data. Alternatively, the model can employ external computational tools, such as a calculator, to ensure precise numerical accuracy. The choice between these approaches can lead to discrepancies, particularly when the model’s internal heuristics produce an answer that deviates from the exact computation provided by an external tool.

\paragraph{Example 2: Factual Data Retrieval}
An LLM responding to fact-based queries may generate responses based on its internalized knowledge, which is constrained by the temporal limitations of its training data. Conversely, it may retrieve real-time information from external databases or search engines. Discrepancies emerge when the internally stored knowledge contradicts newly updated facts, leading to potential inconsistencies in responses.

\paragraph{Example 3: Code Execution}
For programming-related inquiries, an LLM may either generate code snippets based on its learned distribution of syntax and semantics or execute the code using an external interpreter. If the generated code contains errors or outdated syntax, whereas the executed code produces a different, verifiable output, conflicts may arise in determining which response is most reliable.

\paragraph{Example 4: Medical Diagnosis}
In medical applications, an LLM may generate responses based on training data comprising past medical literature and case studies. Alternatively, it may leverage external medical databases containing the latest research findings, treatment protocols, and clinical guidelines. Conflicts may arise when a model’s outdated medical knowledge contradicts contemporary best practices or real-time patient data.

In each of these examples, the underlying issue revolves around the reconciliation of an LLM’s static memory with its dynamic ability to engage external tools. Understanding and addressing these conflicts is crucial to ensuring reliability, accuracy, and transparency in AI-assisted decision-making systems.

\subsection{Causes of Tool-Memory Conflict}
The core of tool-memory conflict stems from a discrepancy between tool-based and parametric knowledge. We identify several causes of tool-memory conflicts.

\paragraph{Temporal Information Mismatch}
Discrepancies occur when tools provide outdated, inconsistent, or delayed information compared to the LLM's internal knowledge. Since LLMs rely on a combination of learned and real-time tool-based data, mismatches can create confusion when attempting to synthesize responses~\citep{jang-etal-2022-temporalwiki}.

\paragraph{Misinformation by Tools}
Errors arise when tools return incorrect, biased, or misleading data, leading to conflicts with the model's stored knowledge. These issues can stem from tool limitations, external data inaccuracies, or adversarial manipulation of information sources, making it difficult for the LLM to discern reliable information.

\paragraph{Incorrect Tool Usage}
Although an LLM may recognize the need for external tools, it can misuse them by selecting an inappropriate tool, misinterpreting output, or passing incorrect arguments. This can result in incomplete or erroneous task execution, especially when tool documentation is ambiguous or when the model's interpretation of tool parameters is flawed.

\subsection{Difference to Existing Knowledge Conflicts in LLMs} \label{sec:diff}

Though look similar, tool-memory conflict is different from existing knowledge conflicts, especially context-memory conflict~\citep{lee-etal-2022-plug,zhou-etal-2023-context,tan-etal-2024-blinded} and inter-context conflict~\citep{zhang-choi-2021-situatedqa,kasai2023realtime,wan-etal-2024-evidence}.

\paragraph{Differences in Knowledge Representation}
Contextual knowledge, such as in-context examples and retrieved documents from RAG, is processed as input tokens. The LLM treats it like any prompt, embedding it into the current context and influencing token-by-token generation. However, it is constrained by the context window and fades unless explicitly reintroduced.

In contrast, tool knowledge, such as API calls or external computations, is acquired on demand. The model does not "know" the result until execution. Tool outputs are typically appended or referenced post-execution rather than embedded in the token sequence. The model does not "remember" these results unless explicitly stored or re-fed into the prompt. Thus, while contextual knowledge is integrated into the model’s generation sequence, tool knowledge operates externally.

\paragraph{Information Flow and Processing Pipeline}
Retrieved contextual knowledge is fed into the transformer architecture before token generation, shaping the model’s latent representation. Tool knowledge, however, is injected mid-process—retrieved results are incorporated after initial processing, often as separate entities. Unlike contextual knowledge, tool outputs bypass the attention mechanism unless explicitly reintroduced.

\paragraph{Epistemological and Practical Implications}
Contextual knowledge follows a retrieval-and-reasoning paradigm, allowing reinterpretation across prompts. Tool knowledge, however, is authoritative—once executed (e.g., querying a database), the model does not reassess multiple sources.

\paragraph{Time-Sensitivity and Dynamism}
Contextual knowledge is static at retrieval, only updating when a new query is issued. Tool knowledge, however, is dynamic and real-time, fetching the latest data upon each call (e.g., live weather updates or stock prices).

\paragraph{Error Handling}
For contextual knowledge, the model can analyze inconsistencies and adjust responses. With tool knowledge, however, the model must accept outputs as-is, reinforcing the idea that tool knowledge functions as an externalized truth source, distinct from contextual knowledge that the model internalizes and processes differently.





\subsection{Importance}\label{sec:app}
Understanding the knowledge conflict between LLM memory and external tools is essential from various aspects.

\paragraph{Identifying LLM Limitations}
LLMs are constrained by training data limitations, outdated knowledge, and biases. Analyzing conflicts helps pinpoint deficiencies, guiding improvements in model design, dataset curation, and fine-tuning.

\paragraph{Assessing Tool Reliability}
External tools vary in accuracy and trustworthiness. Conflicts with LLM responses highlight potential misinformation, enabling better fact-checking, source prioritization, and AI-assisted decision-making.


\subsection{Extracting Tool-Memory Conflict}

\paragraph{Eliciting Tool-Memory Conflict}

For a given query, we prompt the LLM twice, restricting the LLM to only use its internal parametric knowledge (memory) and to only use external tools respectively. To make sure the LLMs are indeed only using memory or tools as we desired, we add keywords in the prompt such as ``only using your internal memory / external tools'', an approach similar to \citet{xie2024adaptive,wang2024resolving}. 
Additionally, we ask the LLMs to document the process of using tools, including what and how tools are used. We exclude all responses where LLMs fail to follow these instructions of solely using tools or memory.

\paragraph{Identifying Bias of Tool-Augmented LLMs}
Given both output based on internal memory and external tools, we prompt the LLM to resolve the conflict by itself, shown in Figure~\ref{fig:mct} (b). 
This can reveal the bias and tendency of LLMs towards specific type of knowledge on different tasks. Following \citet{wu2024clasheval}, we define memory bias and tool bias as
\begin{itemize}
    \item Memory Bias = $Pr[\texttt{LLM}(q|t) = 0 | \texttt{LLM}(q|t; T) = 1, \texttt{LLM}(q|t) = 0]$ measures the probability the model uses the memory while the tool-based output is correct.
    \item Tool Bias = $Pr[\texttt{LLM}(q|t; T) = 0 | \texttt{LLM}(q|t; T) = 0, \texttt{LLM}(q|t) = 1]$ measures the probability the model prefers tool-based output while the memory is correct.
\end{itemize}


\paragraph{Resolving conflicts}
We also evaluate whether existing knowledge resolving methods can resolve tool-memory conflict.
Opinion-based prompting~\citep{zhou-etal-2023-context} reformulates the input query as opinion-based prompting, which demonstrates strong performance to alleviate inter-context conflicts.
Vigilant prompting~\citep{pan-etal-2023-risk} instructs the LLM to beware of potential misinformation, which is shown to significantly alleviate misinformation and inter-context conflicts.
We also examine if incorporating additional sources of information using Retrieval-Augmented Generation (RAG)~\citep{lewis2020retrieval} can alleviate knowledge conflict.

\section{Experimental Setup} \label{sec:experiment}

\paragraph{Datasets}
We evaluate TMC on the following widely-used benchmarks: 
MMLU~\citep{hendrycks2021measuring},
GSM8K~\citep{cobbe2021training},
MATH-500~\citep{lightman2023lets},
AIME 2024~\citep{aime_2024},
GPQA Diamond~\citep{rein2024gpqa}.
These benchmarks cover a diverse range of tasks, including STEM, humanities \& social science, and long-tail world knowledge.

\paragraph{LLMs}
We conduct the experiments across a wide range of LLMs which are capable of calling external tools, 
1) GPT-4o~\citep{hurst2024gpt4o}, 
2) DeepSeek-V3~\citep{liu2024deepseek}, 
3) LLaMA-3 (3.3 70B, 3.1 8B)~\citep{grattafiori2024llama}, 
4) QWen-2.5 (72B)~\citep{yang2024qwen2},
5) QwQ (32B)~\citep{qwq},
6) Groq-LLaMA-3 8B~\citep{groq_llama},
7) Watt (8B)~\citep{watt_ai}.

\begin{table*}[t]
\begin{center}
\caption{Proportion of tool-memory conflict (TMC) across all evaluated tasks of LLMs.}
\label{tab:tmc_prob}
\begin{tabular}{l|ccc|cccc}
\toprule
 Model &  \multicolumn{3}{c|}{No Conflict} & \multicolumn{4}{c}{Conflict} \\
        & Total & Both=1 & Both=0 & Total & Tool=1 & Mem=1 & Both=0 \\ 
\midrule
GPT-4o          & 83.6 & 74.2 &  9.4 & 14.1 &  7.2 &  2.7 &  4.2 \\
DeepSeek-v3     & 80.5 & 72.7 &  7.8 & 15.3 &  6.2 &  3.3 &  5.8 \\
LLAMA-3.3 70B   & 84.5 & 72.2 & 12.3 & 15.5 &  9.1 &  3.0 &  3.4 \\
QWen-2.5 72B    & 73.1 & 62.3 & 10.8 & 26.9 &  8.3 & 11.6 &  7.0 \\
QwQ             & 24.6 & 21.1 &  3.4 & 75.4 &  5.6 & 33.4 & 36.4 \\
Groq-LLAMA-3 8B & 16.8 & 10.7 &  6.1 & 83.2 &  1.2 & 38.1 & 44.0 \\
Watt 8B         & 51.4 & 34.3 & 17.1 & 48.6 &  4.8 & 18.9 & 24.9 \\
\bottomrule
\end{tabular}
\end{center}
\end{table*}

\begin{figure*}[t]
\begin{center}
\includegraphics[width=0.99\linewidth]{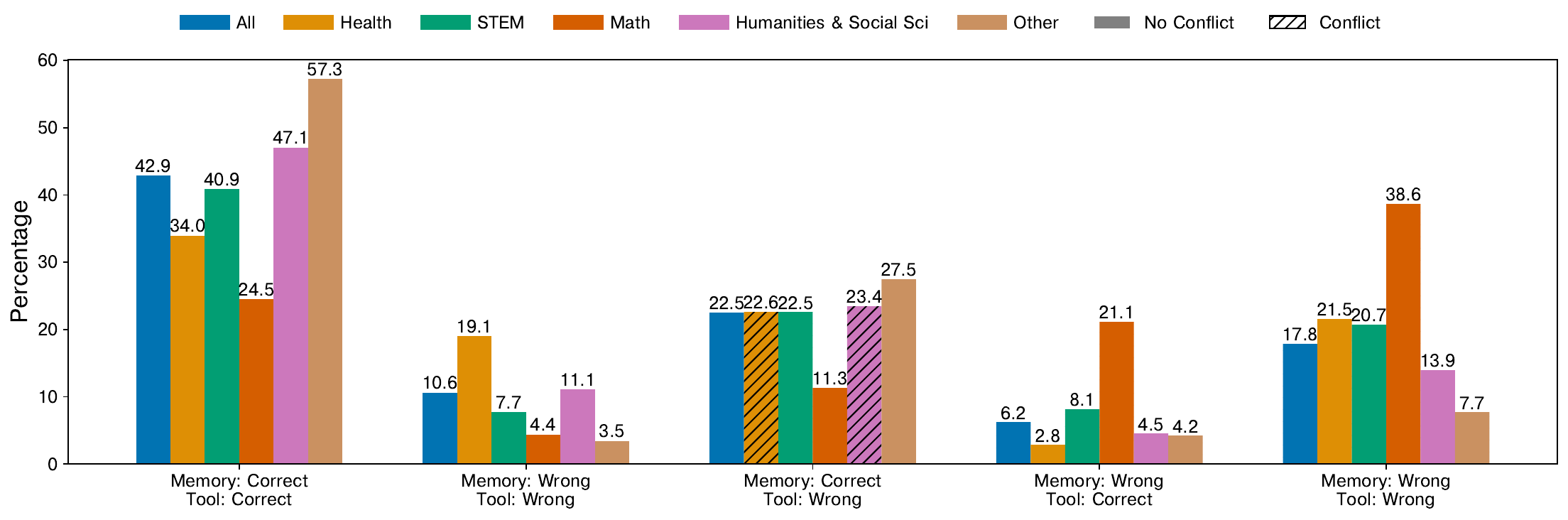}
\end{center}
\caption{Tool-Memory Conflict across different domains.}
\label{fig:group_all}
\end{figure*}

\begin{figure*}[t]
\begin{center}
\includegraphics[width=0.99\linewidth]{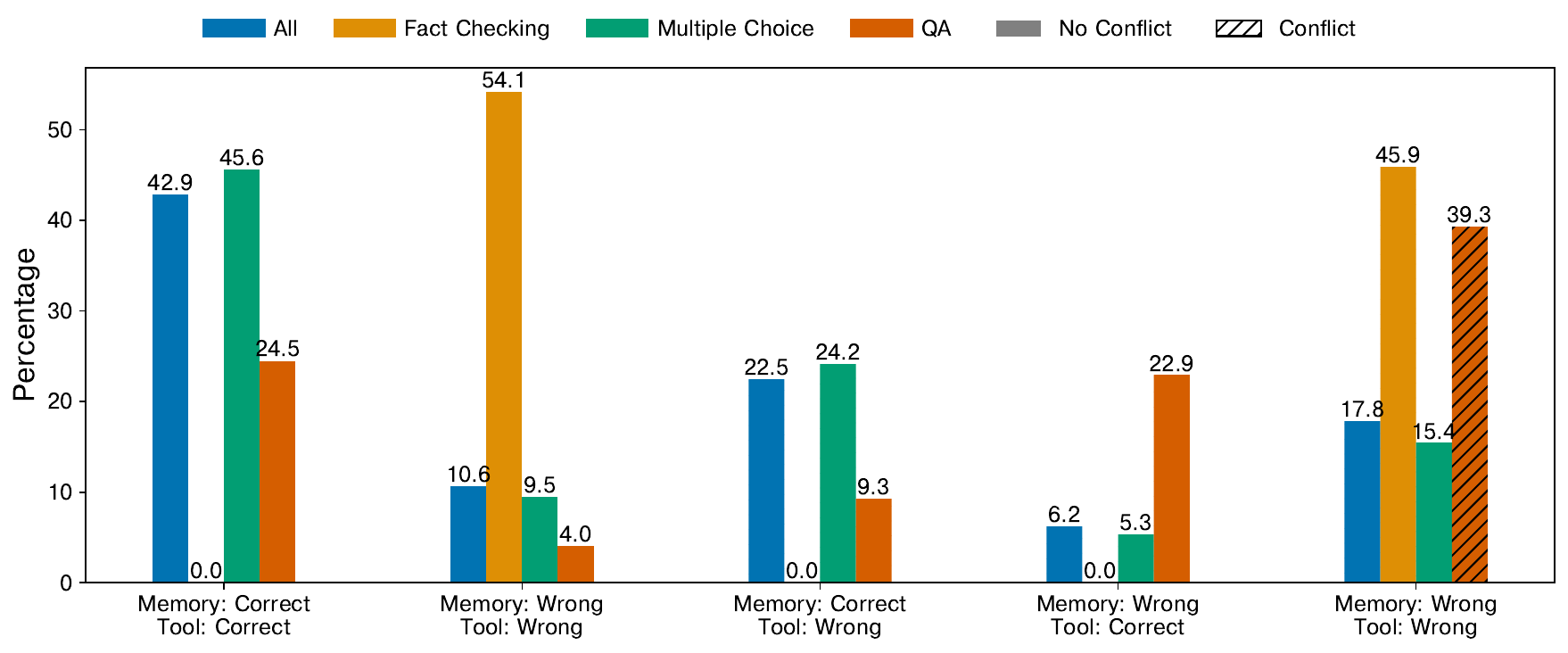}
\end{center}
\caption{Tool-Memory Conflict across different types of tasks.}
\label{fig:task_type_all}
\end{figure*}

\section{Results}

\subsection{Under What Conditions Do Tool-Memory Conflict Occur?}
\paragraph{Prevalence of Tool–Memory Conflict Across Models and Tasks}
Our comprehensive evaluation across seven state-of-the-art LLMs and a broad spectrum of tasks reveals that Tool–Memory Conflict (TMC) is pervasive. On average, 49.8\% of all test instances exhibit a discrepancy between the model’s internally generated answer and the result obtained from an external tool. Specifically, as shown in Table~\ref{tab:tmc_prob}, GPT-4o demonstrates a TMC rate of 14.1\%, DeepSeek-v3 records 15.3\%, LLAMA-3.3 70B yields 15.5\%, QWen-2.5 72B registers 26.9\%, QwQ exhibits a pronounced conflict rate of 75.4\%, Groq-LLAMA-3 8B shows 83.2\%, and Watt 8B records 48.6\%. These figures indicate that even the most advanced LLMs frequently produce outputs that clash with external tool responses. The high conflict rates underscore a fundamental challenge in integrating retrieval or computation tools: the model’s internal knowledge often diverges from—and sometimes directly contradicts—the external source.\looseness-1

\paragraph{Effect of Model Scale on Conflict Frequency}
We observe a clear correlation between model size and TMC incidence. In particular, 70B parameters appear to mark a threshold above which models exhibit substantially lower conflict rates. For example, LLAMA-3.3 70B (15.5\%) and GPT-4o (14.1\%) both outperform their smaller counterparts, Groq-LLAMA-3 8B (83.2\%) and Watt 8B (48.6\%), by wide margins. This suggests that larger models possess richer internal representations, enabling better alignment with external tool outputs. Below the 70B parameter range, LLMs rely more heavily on memorized approximations and heuristics, which are vulnerable to divergence when precision is required or when the tool’s result contradicts the learned distribution.\looseness-1

\paragraph{Domain-Specific Sensitivities to Tool–Memory Conflict}
Figure~\ref{fig:group_all} illustrates how TMC varies across different domains. Domains such as \textit{Math} and \textit{STEM} exhibit the highest conflict probability, indicating that tasks requiring precise numerical computation or specialized technical knowledge are particularly susceptible to discrepancies. In contrast, \textit{Humanities \& Social Sciences} and \textit{Other} tasks show much lower conflict rates, suggesting that, for interpretive or open-ended tasks, the model’s internal reasoning often remains consistent with external cues or that the task does not demand exactitude. \textit{Long-Tail Knowledge} tasks—questions about obscure or infrequently discussed entities—also show moderate TMC levels, likely reflecting mismatches between the model’s outdated or incomplete training data and the tool’s up-to-date information.

\paragraph{Impact of Conflict on Model Confidence and Accuracy}
Whenever a conflict arises, the model must reconcile competing signals: its internal knowledge and the tool’s output. We measure the downstream effect on task performance by comparing accuracy under “No Conflict” versus “Conflict” conditions. In \textit{Math} tasks, accuracy drops by an average of 4.5 absolute points when a tool–memory conflict occurs, indicating that numerical discrepancies severely undermine the model’s confidence and ability to produce correct solutions. For \textit{STEM} and \textit{Health} tasks, we observe a smaller but still notable decline, reflecting the dependence on precise or domain-specific knowledge that external tools often supply. By contrast, \textit{Humanities \& Social Sciences} tasks see only a 0.5 difference between “No Conflict” and “Conflict” conditions, suggesting that models can often override or reinterpret minor contradictions when dealing with more subjective or contextual content. Even for \textit{Long-Tail Knowledge}, the accuracy gap remains modest, implying that models and tools occasionally share similar erroneous assumptions, thereby muting the overall effect of conflict.

\paragraph{Task Type Analysis: Fine-Grained View of Conflict}
Figure~\ref{fig:task_type_all} provides a breakdown of TMC by task type (e.g., arithmetic, algebra, logical reasoning, multi-hop retrieval, etc.). \textit{Arithmetic} and \textit{Algorithmic} tasks exhibit the highest conflict frequencies (often exceeding 70–80\%), underscoring that deterministic, step-by-step computations are prone to mismatch when LLMs rely on learned heuristics rather than exact algorithms. In \textit{Multi-Hop Retrieval} tasks—where the model must chain multiple facts—TMC rates hover around 50–60\%, indicating that each retrieval step compounds the chance of divergence. \textit{Fact Verification} tasks, which require confirming or refuting a given statement, show lower conflict rates, implying that model predictions and tool retrievals tend to align more often. For \textit{Common-Sense Reasoning} tasks, conflict rates fall in the 30–40\% range, suggesting that while external knowledge occasionally contradicts the model’s intuition, the impact is less severe than in strictly quantitative tasks.

\paragraph{Conflict Cases Where Both Memory and Tool Are Incorrect}
A nontrivial fraction of conflict instances occur when both the model’s internal response and the external tool’s output are incorrect but disagree with each other. These “Both=0” cases (Table~\ref{tab:tmc_prob}, last column under “Conflict”) illustrate scenarios in which neither internal memory nor the external tool holds the correct answer. For example, on QwQ, 36.4\% of all test cases fall into this category; for Groq-LLAMA-3 8B, 44.0\% are “Both=0.” Such phenomena highlight that simply combining two erroneous sources does not guarantee improved performance, and that conflict detection alone does not resolve fundamental knowledge gaps.

\paragraph{Mathematical Tasks Exacerbate Tool–Memory Conflicts}
TMC is markedly more pronounced for strictly mathematical tasks. Table~\ref{tab:tmc_prob} shows that, for arithmetic problems, the majority of models exhibit conflict rates well above 70\%. This is attributable to the inherently precise nature of mathematics: external calculators or symbolic solvers provide exact results, whereas LLMs often produce approximate answers based on statistical patterns. Consequently, whenever an LLM’s heuristic estimate deviates by even a small margin, the conflict is detected. Moreover, the discrepancy is magnified in multi-step problems (e.g., solving systems of equations), where compounding rounding errors or misapplied rules lead to large divergences from the tool’s output.

\paragraph{Tool–Memory Conflict Exposes Limitations of Tool-Augmented LLMs}
One of the primary motivations for augmenting LLMs with external tools is to compensate for gaps in the model’s internal knowledge—particularly for long-tail or recently emerged entities. However, our analysis in the \textit{Long-Tail World Knowledge} domain reveals that, even when a tool retrieves the correct answer, the LLM’s internal memory often contains similar inaccuracies. As a result, conflict does not always correspond to a scenario where the tool corrects the model; rather, both sources can share the same outdated or erroneous information. In Figure~\ref{fig:group_all}, the modest difference between “Tool=1” and “Mem=1” under Long-Tail conditions indicates that neither memory nor tool consistently provides a clear advantage. This underscores that tool-augmentation alone cannot rectify deep-seated knowledge deficiencies without also addressing the quality of the underlying retrieval mechanism and ensuring that the external knowledge base is regularly updated.

\paragraph{Domain-Aware Conflict Mitigation Strategies}
Given the varying sensitivity to TMC across domains, we propose that conflict mitigation should be domain-specific. For \textit{Math} and \textit{STEM} tasks, integrating high-precision computational modules (e.g., symbolic solvers, exact arithmetic engines) directly into the LLM’s inference pipeline can drastically reduce discrepancies. In \textit{Humanities \& Social Sciences}, where conflicts are rare and often inconsequential, a simple “favor memory” or “favor tool” heuristic may suffice. For \textit{Long-Tail Knowledge}, a hybrid retrieval strategy that cross-verifies multiple external sources before presenting an answer could minimize the propagation of outdated information. Ultimately, these domain-aware approaches seek to reconcile internal representations with external data in a more principled manner, rather than relying on a one-size-fits-all fallback.

These insights pave the way for future work aimed at harmonizing internal model knowledge with external tool outputs, thereby enhancing the reliability and accuracy of tool-augmented language systems.

\begin{table}[t]
\begin{center}
\caption{Tool Bias and Memory Bias of LLMs.}
\label{tab:bias}
\begin{tabular}{l|c|c}
\toprule
Model & Tool Bias & Memory Bias \\ 
\midrule
GPT-4o            & 41.7 & 41.9 \\
DeepSeek-v3       & 39.2 & 41.3 \\
LLAMA-3.3 70B     & 44.3 & 40.2 \\
QWen-2.5 72B      & 35.8 & 37.3 \\
QwQ               &  0.1 & 24.5 \\
Groq-LLAMA-3 8B   &  0.2 & 16.4 \\
Watt 8B           &  0.0 & 51.4 \\
\bottomrule
\end{tabular}
\end{center}
\end{table}


\subsection{Are LLMs Biased Towards Tools or Memory?}

\paragraph{Tool Bias}
Tool bias is defined here as the fraction of outputs in which the model overly relies on an external tool (for instance, invoking or favoring API-driven lookup routines) instead of integrating information directly. LLAMA-3.3 70B exhibits the highest tool bias at 44.3\%, suggesting that in nearly half of its responses, it defaults to leveraging the attached tool rather than grounding its output solely in internal reasoning. GPT-4o and DeepSeek-v3 have intermediate tool biases of 41.7\% and 39.2\%, respectively. QWen-2.5 72B manifests a lower degree of tool dependence (35.8\%). The three lower-accuracy models—QwQ (0.1\%), Groq-LLAMA-3 8B (0.2\%), and Watt 8B (0.0\%)—demonstrate virtually no tool bias, indicating that they almost never defer to external tools and instead rely exclusively on internal parameters (even though this may come at the cost of accuracy). See Table~\ref{tab:bias} for details.

\paragraph{Memory Bias}
Memory bias represents the probability of and LLM prioritizing where the model disproportionately leverages cached internal knowledge (e.g., memorized facts, pretrained tokens) over dynamic reasoning or tool-assisted querying. GPT-4o and DeepSeek-v3 report memory biases of 41.9\% and 41.3\%, respectively, closely paralleling their tool biases. LLAMA-3.3 70B incurs a slightly lower memory bias (40.2\%), indicating that its reliance on internal memory is marginally less pronounced than its inclination toward tool usage. QWen-2.5 72B’s memory bias (37.3\%) is also lower, reflecting a more balanced distribution between tool usage and memory recall. Conversely, QwQ and Groq-LLAMA-3 8B have memory biases of 24.5\% and 16.4\%, respectively, which—when combined with their negligible tool bias—suggests that these smaller models tend to produce outputs rooted almost entirely in internal knowledge, albeit with far lower overall correctness. Watt 8B exhibits a memory bias of 51.4\%, which exactly matches its overall accuracy score; this suggests that whenever Watt 8B answers correctly, it does so purely via internal recall, never invoking external tools. See Table~\ref{tab:bias} for details.

\paragraph{Comparisons across LLMs}
The two highest-performing models (LLAMA-3.3 70B and GPT-4o) achieve accuracy rates above 83\%, but they differ subtly in how they allocate “cognitive” effort between external tool calls and internal memory. LLAMA-3.3 70B slightly favors the tool (44.3\% tool bias vs. 40.2\% memory bias), whereas GPT-4o displays nearly equal reliance on tools (41.7\%) and memory (41.9\%), reflecting a more balanced hybrid strategy.

DeepSeek-v3 (80.5\% accuracy) also exhibits a nearly balanced tool/memory split (39.2\% vs. 41.3\%), suggesting that its architectural design equally privileges pretrained knowledge and external lookups to achieve high performance. In contrast, QWen-2.5 72B, despite having a 72-billion-parameter backbone, achieves only moderate accuracy (73.1\%) and shows a moderately lower dependence on both tools and memory, implying potential architectural or training differences that reduce over-reliance on either resource.

Smaller models (QwQ, Groq-LLAMA-3 8B, Watt 8B) uniformly demonstrate low tool bias (less than 1.0\%), indicating they are effectively “tool-agnostic.” Their primary—and sometimes sole—decision driver is memorized knowledge, as evidenced by their nonzero memory biases. However, the trade-off is clear: these models achieve sub-optimal accuracy (ranging from 16.8\% to 51.4\%). Watt 8B is the only low-capacity model that manages to achieve accuracies above 50\%, but it does so entirely through memorized content, with no tool assistance.

The close correspondence between accuracy and memory bias in Watt 8B, along with the negligible tool bias, illustrates a scenario in which model capacity is sufficient to store a limited subset of facts (achieving correct responses on roughly half of the dataset) but insufficient to generalize beyond static memorization or effectively integrate external tools. On the other hand, models such as LLAMA-3.3 70B and GPT-4o achieve higher task coverage by intelligently deciding when to call upon external tools versus internal representations, thereby balancing recall and dynamic retrieval.

\begin{table}[t]
\begin{center}
\caption{Resolving Tool-Memory Conflicts}
\label{tab:resolve}
\begin{tabular}{p{2cm}|p{1cm}p{1cm}p{1cm}p{1cm}}
\toprule
 Model & Conflict & Vig Prompt & Op Prompt & RAG \\
\midrule
GPT-4o          & 14.1 & 13.8 & 14.7 & 11.5 \\
DeepSeek-v3     & 15.3 & 14.6 & 15.3 & 12.2 \\
LLAMA-3.3 70B   & 15.5 & 14.4 & 16.3 & 13.6 \\
QWen-2.5 72B    & 26.9 & 22.6 & 25.1 & 17.9 \\
QwQ             & 75.4 & 71.7 & 69.3 & 55.7 \\
Groq-LLAMA-3 8B & 83.2 & 81.5 & 82.2 & 74.3 \\
Watt 8B         & 48.6 & 44.7 & 46.1 & 39.1 \\
\bottomrule
\end{tabular}
\end{center}
\end{table}

\subsection{Can Tool-Memory Conflicts Be Resolved?}


Table~\ref{tab:resolve} provides a quantitative comparison of seven large language models (LLMs) on their propensity to exhibit “tool‐memory conflicts” under different mitigation strategies. In each row, the first column lists a particular LLM, ordered roughly from highest‐capacity (GPT-4o, LLAMA-3.3 70B) to more compact configurations (Groq-LLAMA-3 8B, Watt 8B). The subsequent four columns report the measured conflict rate—expressed as a percentage—under four distinct configurations:

\begin{itemize}
    \item Conflict: This is the raw conflict probability when the model is allowed to choose freely between relying on its internal memorized knowledge and invoking an external tool.
    \item Vigilant prompting: In this case, specialized “vigilance” wording in the prompt is prepended to each query, explicitly instructing the model to detect and avoid contradictory signals between its memory and the tool’s output.
    \item Opinion-based prompting: Opinion-based prompting reformulates the output of LLMs as someone's opinion.
    \item RAG: Additional informaiton is retrieved from external knowledge source and incorporated to answer the query. The model is thus expected to defer to retrieved facts, thereby reducing contradictory reliance on stale internal parameters.
\end{itemize}

\paragraph{Impact of prompt engineering}
Across all architectures, appending a vigilance prompt consistently reduces conflict by approximately 1–4 percentage points relative to the baseline. This suggests that explicit meta‐instructions—e.g., “If your internal knowledge conflicts with the tool’s information, defer to the tool”—do encourage better alignment. However, the magnitude of improvement is modest for the smallest models, which likely lack the representational capacity to fully internalize the meta‐instruction. 

For opinion-based prompting, the conflict probability that either match or slightly exceed the baseline—particularly in LLAMA-3.3 70B and QWen-2.5 72B—implying that not every prompt‐based or fine‐tuning approach uniformly aids consistency. Without explicit documentation, one might tentatively infer that the third intervention either over‐restricts the model (e.g., too aggressive a filter) or fails to provide enough context to override entrenched internal biases.

\paragraph{Effectiveness of RAG}:
Incorporating a retrieval stage proves to be the most effective single intervention: on average, RAG reduces conflict rates by 2–6 percentage points in high‐capacity models and by 8–15 percentage points in mid‐ and low‐capacity models. This indicates that augmenting model “context” with up‐to‐date, externally retrieved evidence not only enriches factual accuracy but also resolves contradictions between stale memorization and the most current tool output. In other words, RAG effectively “grounds” the model’s predictions, irrespective of its size.

\paragraph{Scaling effects on model consistency}
The baseline conflict rates correlate inversely with model size. Larger models (GPT-4o, LLAMA-3.3 70B, DeepSeek-v3) demonstrate stronger internal coherence between memorized knowledge and the outputs of any connected tool. By contrast, smaller models often encode more fragmented or less‐robust knowledge representations, leading to pronounced tool‐memory mismatch.

These findings suggest that more principled approaches, potentially involving model fine-tuning or architectural adjustments, may be necessary to robustly resolve TMC without sacrificing performance or coherence.

\section{Conclusion}
We propose a new type of knowledge conflicts for LLMs -- Tool-Memory Conflict (TMC), where external tools contradicts the internal parametric knowledge of LLMs. Through experiments on diverse datasets, we find that existing LLMs, including powerful proprietary models, suffer from TMC, especially in STEM tasks. Under different conditions, LLMs may be biased towards internal memory or external tools. We evaluate existing tools of resolving conflicts, where prompting-based methods have limited contribution. Incorporating additional external knowledge, such as RAG, may help alleviating the conflicts.

\paragraph{Limitations and Future Work}
We did not experiment with all LLMs, such as Claude. Future work can extend this analysis to a broader range of LLMs. In addition, many conflicting resolving techniques remain not evaluated. We plan to develop novel mechanisms to resolve tool-memory conflicts.


\section*{Impact Statement}
Our work investigates the trustworthiness of integrating external tools into internal parametric knowledge of tool-augmented Large Language Models (LLMs). We focus on identifying a key issue, where the interal memory and external tools conflict with each other. 
Tool-memory conflict undermines model reliability in several ways: (1) it erodes user trust, since end‐users cannot anticipate whether the LLM will defer to its memorized parameters (which may be outdated or imprecise) or to real‐time tool outputs (which may also be noisy or incomplete); (2) it introduces inconsistency in downstream applications—ranging from automated fact‐checking to decision‐support systems—where conflicting signals can lead to erroneous or unsafe conclusions; and (3) it complicates model evaluation and calibration, since standard accuracy metrics fail to capture the nuanced interplay between static knowledge and dynamic retrieval. By systematically quantifying and characterizing the prevalence of tool–memory contradictions, our work illuminates how even state‐of‐the‐art LLMs can exhibit unpredictable behavior when faced with conflicting evidence. In doing so, we highlight the necessity of developing robust conflict‐resolution mechanisms (e.g., vigilance prompting, retrieval‐augmented grounding, and calibrated confidence scoring) that can reconcile or at least surface divergent sources of information. Addressing this gap is essential not only for improving the factual correctness of LLM responses, but also for ensuring transparent, explainable, and trustworthy deployment in high‐stakes domains such as healthcare, finance, and legal assistance.

\bibliography{reference,anthology}
\bibliographystyle{icml2025}

\newpage
\appendix
\onecolumn


\end{document}